\documentclass[sigconf,prologue,table]{acmart}

\usepackage{multirow}
\usepackage{diagbox}
\usepackage{color}
\usepackage{subfigure}
\usepackage{threeparttable}

\newcommand{\tabincell}[2]{\begin{tabular}{@{}#1@{}}#2\end{tabular}}

\newcommand{\lin}[1]{{\textbf{\color{orange}[Lin: #1]}}}

\newcommand{\tool}{\textsc{SIRA}}
\newcommand{\model}{BERT+Attr-CRF}

\usepackage[framemethod=TikZ]{mdframed}
\usepackage{tcolorbox}
\makeatletter
\newcommand{\mybox}[1]{%
  \setbox0=\hbox{#1}%
  \setlength{\@tempdima}{\dimexpr\wd0+13pt}%
  \begin{tcolorbox}[boxrule=0.5pt, colback=white, arc=4pt,
      left=6pt,right=6pt,top=6pt,bottom=6pt,boxsep=0pt]
    #1
  \end{tcolorbox}
}

\AtBeginDocument{%
  \providecommand\BibTeX{{%
    \normalfont B\kern-0.5em{\scshape i\kern-0.25em b}\kern-0.8em\TeX}}}

\copyrightyear{2022}
\acmYear{2022}
\setcopyright{acmcopyright}
\acmConference[ICSE '22]{44th International Conference on Software Engineering}{May 21--29, 2022}{Pittsburgh, PA, USA}
\acmBooktitle{44th International Conference on Software Engineering (ICSE '22), May 21--29, 2022, Pittsburgh, PA, USA}
\acmPrice{15.00}
\acmDOI{10.1145/3510003.3510189}
\acmISBN{978-1-4503-9221-1/22/05}

\begin{document}

\title{Where is Your App Frustrating Users?}

\author{Yawen Wang$^{1,2}$, Junjie Wang$^{1,2,3,*}$, Hongyu Zhang$^{5}$, Xuran Ming$^{1,2}$, Lin Shi$^{1,2}$, Qing Wang$^{1,2,3,4,*}$}
\thanks{$^{*}$ Corresponding author.}
\affiliation{$^1$ Laboratory for Internet Software Technologies, $^3$ State Key Laboratory of Computer Sciences, $^4$ Science \& Technology on Integrated Infomation System Laboratory, Institute of Software Chinese Academy of Sciences
\city{Beijing}
\country{China}}
\affiliation{$^2$ University of Chinese Academy of Sciences
\city{Beijing}
\country{China}}
\affiliation{$^5$ The University of Newcastle
\city{Callaghan}
\country{Australia}
\\ \{yawen2018, junjie, xuran2020, shilin, wq\}@iscas.ac.cn, hongyu.zhang@newcastle.edu.au
}




\renewcommand{\shortauthors}{Wang, et al.}

\begin{abstract}
User reviews of mobile apps provide a communication channel for developers to perceive user satisfaction.
Many app features that users have problems with are usually expressed by key phrases such as ``upload pictures'', which could be buried in the review texts.
The lack of fine-grained view about problematic features could obscure the developers' 
understanding of where the app is frustrating users, and postpone the improvement of the apps.
Existing pattern-based approaches 
to extract target phrases suffer from low accuracy due to insufficient semantic understanding of the reviews, thus can only summarize the high-level topics/aspects of the reviews.
This paper proposes a semantic-aware, fine-grained app review analysis approach ({\tool}) to extract, cluster, and visualize the problematic features of apps.
The main component of {\tool} is a novel {\model} model for fine-grained problematic feature extraction, which combines textual descriptions and review attributes to better model the semantics of reviews and boost the performance of the traditional BERT-CRF model. {\tool} also clusters the extracted phrases based on their semantic relations and presents a visualization of the summaries. 
Our evaluation on 3,426 reviews from six apps confirms the effectiveness of {\tool} in problematic feature extraction and clustering.
We further conduct an empirical study with {\tool} on 318,534 reviews of 18 popular apps to explore its potential application and examine its usefulness in real-world practice.

\end{abstract}



\maketitle

\section{Introduction}
\label{sec:intro}

Mobile app development has been active for over a decade, generating millions of apps for a wide variety of application domains such as shopping, banking, and social interactions. They have now become indispensable in our daily life. The importance of mobile apps urges the development team to make every endeavor to 
understand users' concerns and improve app quality. 

Users often write reviews of the mobile apps they are using on distribution platforms such as Apple Store and Google Play Store.
These reviews are short texts that can provide valuable information to app developers, such as user experience, bug reports, and enhancement requests \cite{Johann2017SAFE,DiSorbo2016what,guo2020Caspar,man2016experience}.
A good understanding of these reviews can help developers improve app quality and user satisfaction \cite{Gu2015what,Panichella2015how,Khalid2015what}.
However, popular apps may receive a large number of reviews every day.  Therefore, manually reading and analyzing each user review to 
extract useful information
is very time-consuming. 

\begin{figure}[htb]
\centering
\includegraphics[width=\columnwidth]{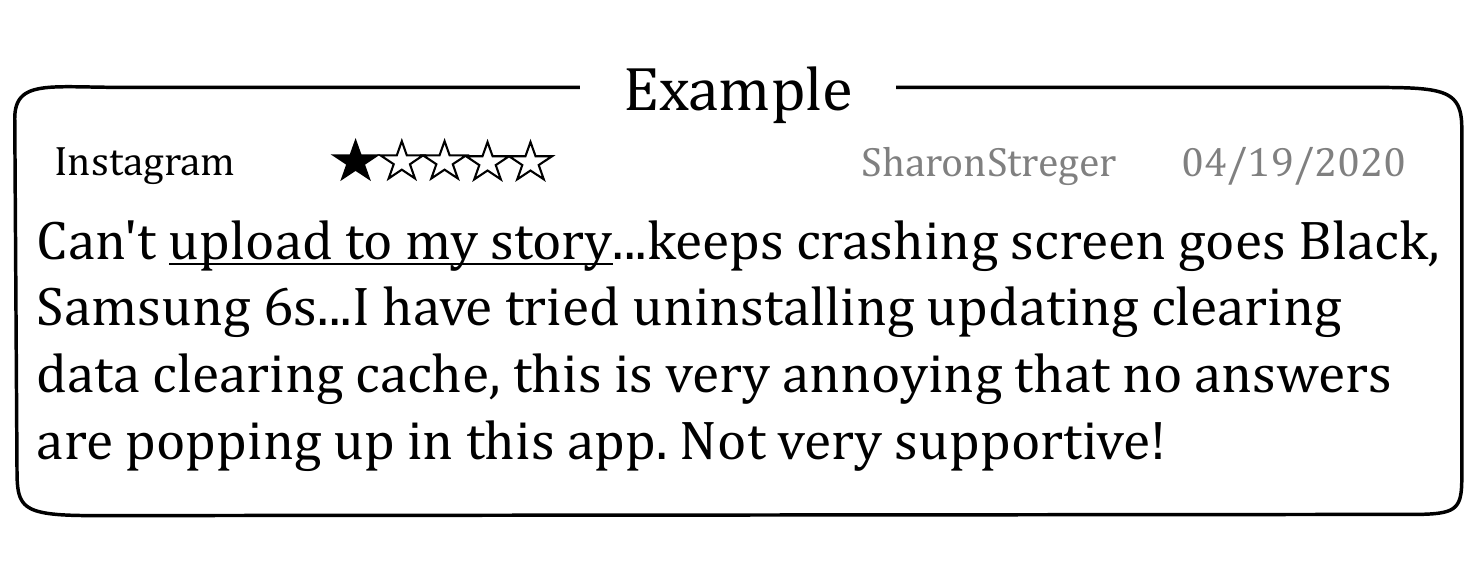}
\caption{An example app review and problematic feature.}
\label{fig:example}
\end{figure}

In recent years, automated techniques for mining app reviews have attracted much attention \cite{harman2012app,palomba2015user,noei2018winning}. These techniques can help reduce the effort required to understand and analyze app reviews in many ways, such as topic discovery \cite{maalej2015bug,Panichella2015how,chen2015ARMiner}, and key phrase extraction \cite{Vu2015mining,Gu2015what,DiSorbo2016what,Johann2017SAFE,gao2018INFAR}.
However, existing work about topic discovery can only identify \textit{WHAT} the users complain about \cite{Khalid2015what,Vu2015mining,Panichella2015how}, such as the high-level topics/aspects of the reviews (e.g., \textit{compatibility},  \textit{update}, \textit{connection}, etc).
Taken the review of \textit{Instagram} in Figure \ref{fig:example} as an example, existing approaches would capture terms such as \textit{update, cache, uninstall}, yet missing its core intent.
Developers still could not have a concrete understanding about which specific features of the app the users are complaining about.
Furthermore, existing work about key phrase extraction mainly utilizes heuristic-based techniques (such as Part-of-Speech patterns, parsing tree, and semantic dependence graph) to extract the target phrases, which could have insufficient semantic understanding of the reviews. As a result, their accuracy is less satisfactory and can be further improved.

In comparison, we aim at exploiting the \textit{WHERE} aspect of the app reviews, and providing an accurate fine-grained landscape about \textit{where an app frustrates the users}, i.e., which specific app features\footnote{We refer to a feature as a distinctive, user-visible characteristic of a mobile app \cite{kangfeature1990}\cite{zhang2006feature}, e.g., sending videos, viewing messages, etc.} the users have problems with.
As an example in Figure \ref{fig:example}, the review is about a crashing problem, and the problematic feature the user complained about is \textit{upload to my story}.
The fine-grained knowledge about problematic features could facilitate app developers in understanding the user concerns, localizing the problematic modules, and conducting follow-up problem-solving activities.

To overcome the drawbacks of existing work and better exploit the app reviews, this paper proposes a Semantic-aware,  
fIne-grained app Review Analysis approach ({\tool}), which can extract, cluster, and visualize the problematic features of apps. 
More specifically, {\tool} includes a novel {\model} model to automatically extract the fine-grained phrases (i.e., problematic features). 
It combines the review descriptions and review attributes (i.e., app category and review description sentiment) to better model the semantics of reviews and boost the performance of the traditional BERT-CRF model \cite{xu2020cluener2020}.
With the extracted phrases, {\tool} then designs a graph-based clustering method to summarize the common aspects of problematic features based on their semantic relations.
Finally, {\tool} presents a visualization of the summarized problematic features. 

We evaluate {\tool} on 3,426 reviews involving 8,788 textual sentences from six apps spanning three categories.
For problematic feature extraction, the overall precision and recall achieved by {\tool} is 84.27\% and 85.06\% respectively, significantly outperforming the state-of-the-art methods. 
{\tool} can also achieve high performance in problematic feature clustering, outperforming two commonly-used baselines.
We further conduct an empirical study with {\tool} on 318,534 reviews of 18 popular apps (reviews spanning 10 months) to explore its potential application and examine its usefulness in real-world practice.
We find that different apps have their unique problematic features and problematic feature distributions. 
The results also reveal that different apps can share some common problematic features. 
This observation can facilitate mobile app testing, e.g., recommending bug-prone features to similar apps for test prioritization.

The main contributions of this paper are as follows:
\begin{itemize}
    \item A semantic-aware, fine-grained app review analysis approach ({\tool}) to extracting, clustering, and visualizing the problematic features of apps.
    In {\tool}, we design a {\model} model to automatically extract the fine-grained phrases (i.e., problematic features), and a graph-based clustering method to summarize the common aspects of problematic features.
    \item The evaluation of the proposed {\tool} on 3,426 reviews involving 8,788 textual sentences from six apps spanning three categories, with affirmative results.
    \item A large-scale empirical study on 318,534 reviews of 18 popular apps, to explore its potential application and usefulness in real-world practice.
    \item Public accessible source code and experimental data at \url{https://github.com/MeloFancy/SIRA}.
\end{itemize}

\section{Background and Related Work}
\label{sec:background}




\textbf{Named Entity Recognition (NER).} NER is a classic Natural Language Processing (NLP) task of sequence tagging \cite{ZhangCZLY18, HuangXY15}.
Given a sequence of words, NER aims to predict whether a word belongs to named entities, e.g., names of people, organizations, locations, etc.
NER task can be solved by linear statistical models, e.g., Maximum Entropy Markov models \cite{McCallumFP00, Tang505}, Hidden Markov Models \cite{oxfordjournals.molbev.a025575} and Conditional Random Fields (CRF) \cite{LaffertyMP01}.
Deep learning-based techniques would use a deep neural network to capture sentence semantics and a CRF layer to learn sentence-level tag rules.
Typical network structures include convolutional neural network with CRF (Conv-CRF) \cite{CollobertWBKKK11}, Long Short-Term Memory network with CRF (LSTM-CRF) and bidirectional LSTM network with CRF (BiLSTM-CRF) \cite{HuangXY15}.
By taking advantage of the bidirectional structure, BiLSTM-CRF model can use the past and future input information and can usually obtain better performance than Conv-CRF and LSTM-CRF.

Language model pre-training techniques have been shown to be effective for improving many NLP tasks \cite{RuderH18, DevlinCLT19}.
BERT (Bidirectional Encoder  Representations  from  Transformers) \cite{DevlinCLT19} is a Transformer-based \cite{VaswaniSPUJGKP17} representation model that uses pre-training to learn from the raw corpus, and fine-tuning on downstream tasks such as the NER task.
Employing BERT to replace BiLSTM (short for BERT-CRF) could lead to further performance boosts \cite{xu2020cluener2020}.
BERT-CRF model benefits from the pre-trained representations on large general corpora combined with fine-tuning techniques.

\begin{figure*}[tb]
\centering
\includegraphics[width=\textwidth]{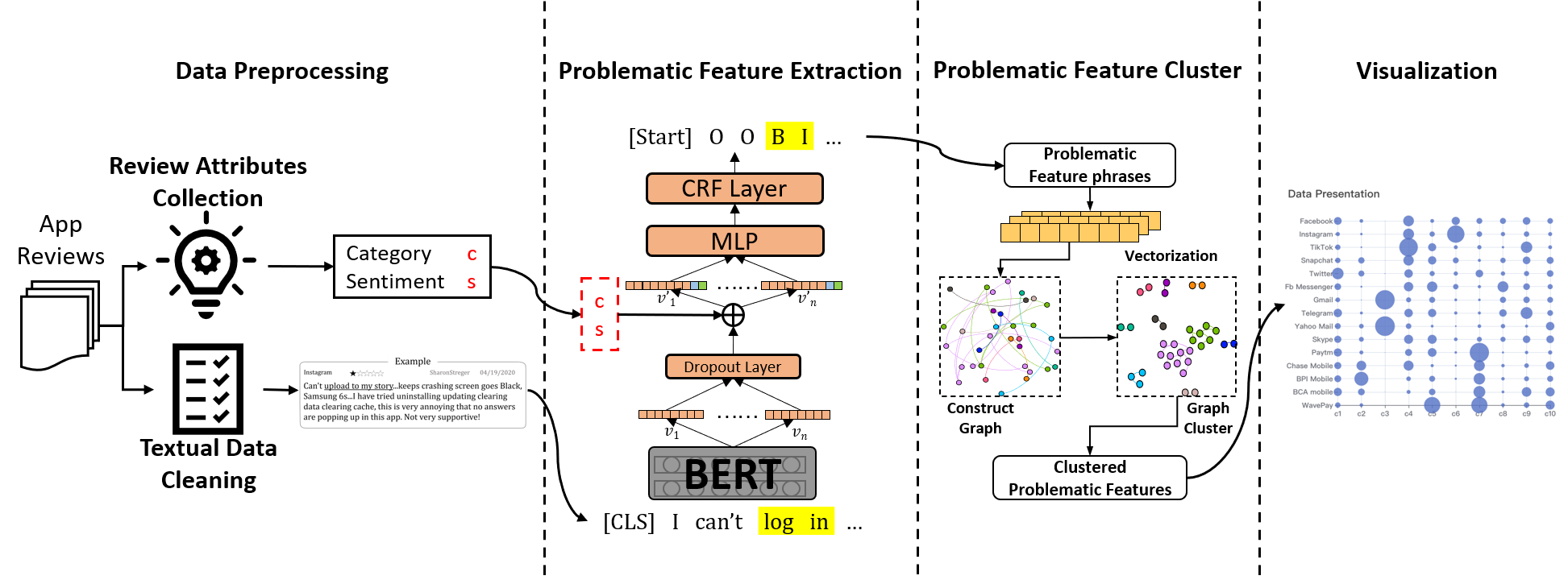}
\caption{The overview of {\tool}.}
\label{fig:overview}
\end{figure*}

\label{sec_related}


\textbf{Mining user reviews.}
Harman et al. introduced the concept of app store mining by identifying correlations between the customer ratings and the download rank of a mobile app \cite{harman2012app,martin2017survey}. 
Palomba et al. found that developers implementing user reviews would be rewarded in terms of app ratings \cite{palomba2015user}.
Noei et al. investigated the evolution of app ranks and identified the variables that share a strong relationship with ranks, e.g., number of releases \cite{noei2018winning}.

Previous studies on mining user reviews emphasized the topic discovery/classification and summarization of reviews as a way of aggregating a large amount of text and reducing the effort required for analysis \cite{maalej2015bug,Panichella2015how,chen2015ARMiner,DiSorbo2016what,OehriG20}.
These classifications are from different points of view, e.g., whether or not the reviews include bug information, requests for new features \cite{maalej2015bug}, whether they are informative \cite{chen2015ARMiner}, whether reviews across different languages and platforms are similar \cite{OehriG20}, or based on a taxonomy relevant to software maintenance and evolution \cite{Panichella2015how}, etc.
Other studies focused on the information extraction from app reviews considering the fact that reading through the entire reviews is impractical \cite{Vu2015mining,Gu2015what,gao2018INFAR,guo2020Caspar,Khalid2015what,KurtanovicM18}.
For example, the types of complains \cite{Khalid2015what}, the app aspects loved by users \cite{Gu2015what}, user rationale \cite{KurtanovicM18} and summaries for guiding release planning \cite{Villarroel2016release} are extracted and summarized for facilitating the review understanding.

There are some studies on mining API-related opinions from informal discussions, such as Q\&A websites (e.g., Stack Overflow) to alleviate developers' burden in performing manual searches \cite{UddinK17a,0008ZBPL19}.
These methods mainly depend on fuzzy matching with pre-built API databases, which cannot work in our context.
There are also some studies on mining social media data (e.g., Twitter data) \cite{GuzmanAS17}.
The app reviews mainly convey users' feedback about an app, while the Twitter data is more general and contains daily messages.
Therefore, general-purpose techniques for Twitter data require customizations to better understand app reviews.

Some studies are similar to our work, such as topic discovery/classification, sentiment analysis, etc.
However, they do not support the extraction of fine-grained features well.
For example, INFAR \cite{gao2018INFAR} mines insights from app reviews and generates summarizes after classifying sentences into pre-defined topics.
The discovered topics from INFAR are more coarse-grained (e.g., GUI, crash, etc.).
Our method can highlight the fine-grained features (e.g., "push notification") that users complained about;
SUR-Miner \cite{Gu2015what} and Caspar \cite{guo2020Caspar} uses techniques, such as dependency parsing and Part-of-Speech pattern, to extract some aspects from app reviews.
Guzman et al. \cite{GuzmanM14} proposed a method, which can only extract features consisting of two words (i.e., collocations) from the reviews based on word co-occurrence patterns, which is not applicable in our context, because the problematic features might contain multiple words;
Opiner \cite{UddinK17a} is a method to mining aspects from API reviews.
It extracts API mentions from API reviews through exact and fuzzy name matching with pre-built API databases, which is difficult to work in our context because we do not have a database of feature phrases in advance.
These studies utilized pattern-based method to extract the target phrases, which did not consider the review semantics sufficiently, and had bad tolerance to noise; by comparison, our proposed approach is a semantic-aware approach.

\textbf{Mining open source bug reports.}
Previous studies have proposed various methods to automatically classify bug reports \cite{liu2020automated,wang2017domain}, detect the duplicate reports \cite{CIKM12, Cooper21ittakes,wang2019images}, summarize the reports \cite{hao2019CTRAS}, and triage the reports \cite{ISSRE14,xia2017improving,lee2017applying}, etc.
The bug reports in open source or crowd testing environment are often submitted by software practitioners, and often described with detailed bug explanation and in relatively longer length.
Yet the app reviews are submitted by the end users and in much fewer words, thus the above mentioned approaches could not be easily adopted in this context. 

\textbf{Semantic-aware approaches in SE.}
Researchers have utilized deep learning based techniques to capture the semantics of software artifacts and facilitate the follow-up software engineering tasks.
Such kinds of studies include neural source code summarization with attentional encoder-decoder model based on code snippets and summaries \cite{zhang2020retrieval}, requirement traceability by incorporating requirements artifact semantics and domain knowledge into the tracing solutions \cite{guo2017semantically}, knowledge mining of informal discussions on social platforms \cite{Wang2020difftech},  etc.
This paper focuses on a different type of software artifact (i.e., app reviews) and incorporates a state-of-the-art technique (i.e., BERT) for the semantic-aware learning, and the results show its effectiveness. 


\section{Approach}
\label{sec:approach}




This paper proposes a Semantic-aware, fIne-grained app Review Analysis approach {\tool} to extract, cluster, and visualize the problematic features of apps (i.e., the phrases in app reviews depicting the feature which users have problems with, see the examples in Figure \ref{fig:example}.)

Figure \ref{fig:overview} presents the overview of {\tool}, which consists of four steps. 
\textbf{\textit{First}}, it preprocesses the app reviews crawled from online app marketplace, to obtain the cleaned review descriptions and the review attributes (i.e., the category of the belonged app $c$ and the review description sentiment $s$). 
\textbf{\textit{Second}}, it builds and trains a {\model} model to automatically extract the fine-grained phrases about problematic features.
{\model} combines the review descriptions and two review attributes as input to better model the semantics of reviews and boost the phrase extraction performance of the traditional BERT-CRF model.
\textbf{\textit{Third}}, {\tool} clusters the extracted phrases with a graph-based clutering method to summarize the common aspects of problematic features based on their semantic relations.
And \textbf{\textit{finally}}, it presents a visualization view to illustrate the summaries and compare the problematic features among apps, in order to acquire a better understanding of where users complain about across apps.


\subsection{Data Preprocessing}
\label{subsec:approach_data_parsing}


Data preprocessing 
mainly includes two steps: textual data cleaning and review attribute collection.

\subsubsection{\textbf{Textual Data Cleaning}}
\
\newline
The raw app reviews are often submitted via mobile devices and typed using limited keyboards.
This situation leads to the frequent occurrences of massive noisy words, such as repetitive words, misspelled words, acronyms and abbreviations \cite{GaoZLK18, Vu2015mining, VuPNN16, Gu2015what}.

Following other CRF-based practices \cite{HuangXY15}, we treat each sentence as an input unit.
We first split each review into sentences by matching punctuations through regular expressions.
Then we filter all non-English sentences with Langid\footnote{https://github.com/saffsd/langid.py}.
We tackle the noisy words problem with the following steps:
\begin{itemize}
    \item \textbf{Lowercase}: we convert all the words in the review descriptions into lowercase.
    \item \textbf{Lemmatization}: we perform lemmatization with Spacy\footnote{https://spacy.io} to alleviate the influence of word morphology.
    \item \textbf{Formatting}: we replace all numbers with a special symbol ``<number>'' to help the BERT model unify its understanding.
    Besides, we build a list containing all the app names crawled from Google Play Store, and replace them with a uniform special symbol ``<appname>''.
\end{itemize}

\subsubsection{\textbf{Review Attribute Collection}}
\label{subsubsec:approach_extract_attr}
\
\newline
Some attributes related to the review or the app 
can facilitate the extraction of problematic features in Section \ref{subsec:approach_phrase_extraction}.
This subsection collects these attributes, i.e., the category of the belonged app $c$ and the review description sentiment $s$ as shown in Figure \ref{fig:overview} and Figure \ref{fig:incorporate_attr}.
The reason why we include the app category is 
that apps from different categories would exert unique nature in terms of functionalities and topics \cite{GaoZX0LK19}. 
Furthermore, review descriptions with negative sentiment would be more likely to contain problematic features, compared with the description with positive sentiment. 
Hence, we include review description sentiment as the second attribute in our model.



App categories can be directly collected when crawling data from Google Play Store.
To obtain the sentiment for each review sentence, we employ SentiStrength-SE \cite{IslamZ18}, a domain-specific sentiment analysis tool especially designed for software engineering text.
SentiStrength-SE would assign a positive integer score in the range of 1 (not positive) to 5 (extremely positive) and a negative integer score in the range of -1 (not negative) to -5 (extremely negative) to each sentence.
Employing two scores is because previous research from psychology \cite{PMID:25926805} has revealed that human beings process the positive and negative sentiment in parallel.
Following previous work \cite{GuzmanM14, GaoZX0LK19}, if the absolute value of the negative score multiplied by 1.5 is larger than the positive score, we assign the sentence the negative sentiment score; otherwise, the sentence is assigned with the positive sentiment score.

\subsection{Problematic Feature Extraction}
\label{subsec:approach_phrase_extraction}

We model the problematic feature extraction problem as a Named Entity Recognition (NER) task, where we treat problematic features as named entities, and solve the problem with the commonly-used CRF technique.
To better capture the semantics of the app reviews, we employ the BERT model to encode the review descriptions.
Furthermore, we incorporate the review attributes in the CRF model to further boost the recognition of problematic features.
Two attributes, i.e., category of the belonged app $c$ and review description sentiment $s$ (see Section \ref{subsubsec:approach_extract_attr}), are utilized in our model.

Following other NER tasks, we use the BIO tag format \cite{RatinovR09, DaiLCT15} to tag each review sentence, where
\begin{itemize}
    \item B-label (Beginning): The word is the beginning of the target phrase.
    \item I-label (Inside): The word is inside the target phrase but not its beginning.
    \item O-label (Outside): The word is outside the target phrase.
\end{itemize}
The BIO-tagged review sentence is input into the {\model} model for further processing.

\begin{figure}[tb]
\centering
\includegraphics[width=0.7\columnwidth]{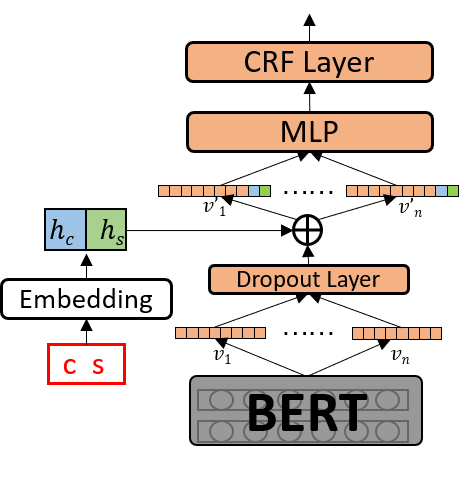}
\caption{Detailed structure of {\model}.}
\label{fig:incorporate_attr}
\end{figure}

Figure \ref{fig:incorporate_attr} presents the detailed structure of our proposed {\model} model.
Since app reviews are short texts, and the involved vocabulary is relatively small,
we use the pre-trained model $BERT_{BASE}$\footnote{https://huggingface.co/bert-base-uncased}, which has 12 layers, 768 hidden dimensions and 12 attention heads.
It has been pre-trained on the BooksCorpus (800M words) and English Wikipedia (2,500M words), and will be fine-tuned using our own data.
Each input sentence is represented by 128 word tokens with a special starting symbol $[CLS]$. 
For those not long enough, we use a special symbol $[PAD]$ to pad them to the length of 128,  following the common practice.
The outputs of BERT are fed into a dropout layer to avoid over-fitting.
Finally, we obtain $n$ (the length of the input sentence) vectors, with each vector (denoted as $v_i$) having 768 dimensions and corresponding to each input word.  

We incorporate the review attributes into the textual vectors ($v$) to jointly capture the underlying meaning of the review sentence.
The review attributes ($c$ and $s$) extracted in Section \ref{subsubsec:approach_extract_attr} are discrete values.
We first convert them into continuous vectors (denoted as $h_c$ and $h_s$) by feeding them into the embedding layers.
Taking attribute $s$ as an example, it can take ten values (-5 to -1 and 1 to 5).
The embedding layer could represent each value with a continuous vector, which can be trained jointly with the whole model.
We then concatenate $h_c$, $h_s$ and $v$ ($h_c \bigoplus h_s \bigoplus v$) to obtain a vector (denoted as $v_i'$) for each input word.
The concatenated vectors first go through a Multi-layer Perceptron (MLP), which computes the probability vector (denoted $p$) of BIO tags for each word:
\begin{equation}
    p=f(W [h_c;h_s;v])    
\end{equation}
where $f(\cdot)$ is the activation function, and $W$ is trainable parameters in MLP.
$[h_c;h_s;v]$ is the concatenation of these three vectors.
Finally, $p$ is input into the CRF layer to determine the most likely tag sequence based on Viterbi Algorithm \cite{viterbi}.

Based on the derived tag sequence, we can obtain the phrases about problematic features. 
For example, if our input review sentence is ``whenever I go to send a video it freezes up'', and the output tag sequence is ``$<O> <O> <O> <O> <B> <I> <I> <O> <O> <O>$'', we can determine the extracted problematic feature as ``send a video'' based on the BIO format.


The loss function of the model should measure the likelihood of the whole true tag sequence, instead of the likelihood of the true tag for each word in the sequence. 
Therefore, the commonly-used Cross Entropy is not suitable in this context. 
Following existing studies \cite{HuangXY15}, the loss function contains two parts: the emission score and the transition score.
It is computed as:
\begin{equation}
    s([x]^T_1,[l]^T_1,\widetilde{\theta})=\sum_{t=1}^{T}([A]_{[l]_{t-1},[l]_t}+[f_\theta]_{[l]_t,t})
\end{equation}
where $[x]^T_1$ is the sentence sequence of length $T$, and $[l]^T_1$ is the tag sequence.
$f_\theta([x]^T_1)$ is the emission score, which is the output of MLP with parameters $\theta$, and $[A]_{i,j}$ is the transition score, which is obtained with the parameters from the CRF layer.
The transition score $[A]_{i,j}$ models the transition from the $i$-th state to the $j$-th state in the CRF layer.
$\widetilde{\theta}=\theta\cup \left \{ [A]_{i,j}\forall i,j \right \}$ is the new parameters for the whole network.
The loss of a sentence $[x]^T_1$ along with a sequence of tags $[l]^T_1$ is derived by the sum of emission scores and transition scores.

\textbf{Model Training:}
The hyper-parameters in {\tool} are tuned carefully with a greedy strategy to obtain the best performance.
Given a hyper-parameter $P$ and its candidate values $\left \{v_1, v_2,...,v_n \right \}$, we perform automated tuning for $n$ iterations, and choose the values which leads to the best performance as the tuned value of $P$.
After tuning, the learning rate is set as $10^{-4}$. The optimizer is Adam algorithm \cite{KingmaB14}. We use the mini-batch technique for speeding up the training process with batch size 32. The drop rate is 0.1, which means 10\% of neuron cells will be randomly masked to avoid over-fitting.

We implement this {\model} model using Transformers\footnote{https://github.com/huggingface/transformers}, which is an open-source Pytorch library for Natural Language Understanding and Natural Language Generation. Our implementation and experimental data are available online\footnote{https://github.com/MeloFancy/SIRA}.


\subsection{Problematic Feature Clustering}
\label{subsec: approach_phrase_topic_cluster}


The extracted problematic features might be linguistically different yet semantically similar.
To provide a summarized view of the problematic features, this step clusters the extracted problematic features based on the topics derived from their semantic relations.
Conventional topic models use statistical techniques (e.g., Gibbs sampling) based on word co-occurrence patterns \cite{PorteousNIASW08}.
They are not suitable for the short texts (i.e., problematic features in our context), because the co-occurrence patterns can hardly be captured from the short text, instead the semantic information should be taken into consideration.
Additionally, these models need to specify the number of clusters/topics, which is hardly determined in our context.
To tackle these challenges, we design a graph-based clustering method, which employs semantic relations of problematic features.



\textbf{First}, we convert problematic feature phrases into 512 dimensional vectors using Universal Sentence Encoder (USE) \cite{CerYKHLJCGYTSK18}.
It is a transformer-based sentence embedding model that captures rich semantic information, and has been proven more effective than traditionally-used word embedding models \cite{guo2020Caspar}.
\textbf{Second}, we construct a weighted, undirected graph, where each problematic feature is taken as a node, and the cosine similarity score between USE vectors of two problematic features is taken as the weight between the nodes.
If the score is over a certain ratio, we add an edge between two nodes. 
The ratio is an input hyper-parameter, which measures the semantic correlations between problematic features. The higher ratio leads to higher cluster cohesion.
We set it as 0.5 after tuning in the training data.
\textbf{Third}, we perform Chinese Whispers (CW) \cite{2006Chinese}, which is an efficient graph clustering algorithm, on this graph to cluster problematic features.

With this graph-based clustering method, {\tool} can group the problematic features that are semantically similar into the same topic.
We implement our clustering method in python, based on the open-source implementation of USE\footnote{https://github.com/MartinoMensio/spacy-universal-sentence-encoder} and CW\footnote{https://github.com/nlpub/chinese-whispers-python}.

\subsection{Visualization}
\label{subsec:approach_visualization}

In order to display the clustering results of multiple apps more intuitively, we provide a visualized view in the form of bubble charts (an example is shown in Figure \ref{fig:cluster_stats}).
The y-axis demonstrates the names of investigated apps, and the x-axis represents the id of each cluster.
The size of the bubble (denoted as $s_{a,c}$) of app $a$ in cluster $c$ is defined as the ratio between the number of problematic features of app $a$ in cluster $c$ and the total number of problematic features in app $a$.

When the cursor hovers over the bubble, it would display detailed information of this cluster, including the cluster name, the number of problematic features, and example reviews with corresponding problematic features.
For the cluster name, we first find the most frequent noun or verb (denoted as $w$) among all problematic features in the cluster.
We then count the number of problematic features containing $w$, and treat the most frequent phrase as the cluster name (i.e., the representative problematic feature).
By comparing the relative sizes of bubbles, one can intuitively acquire the distribution of problematic features across apps.


\section{Experimental Design}
\label{sec:experiment}


\subsection{Research Questions}
\label{subsec:experiment_RQ}

We answer the following three research questions:

\begin{itemize}
\item \textbf{RQ1:} What is the performance of {\tool} in extracting problematic features?

\item \textbf{RQ2:} Is each type of the review attributes employed in {\tool} necessary? 

\item \textbf{RQ3:} What is the performance of {\tool} in clustering problematic features?

\end{itemize}

\textbf{RQ1} investigates the performance of {\tool} in problematic feature extraction, and we also compare the performance with four  
state-of-the-art baselines (see Section \ref{subsec:experiment_baseline}) to further demonstrate its advantage.
\textbf{RQ2} conducts comparison with {\tool}'s three variants to demonstrate the necessity of the employed review attributes in {\model} model.
\textbf{RQ3} investigates the performance of {\tool} in problematic feature clustering, and we also compare {\tool} 
with two commonly-used baselines (see Section \ref{subsec:experiment_baseline}).




\subsection{Data Preparation}
\label{subsec:experiment_data_prepare}

We use the reviews of six apps from three categories (two in each category) in our experiments.
All six apps are popular and widely-used by a large number of users.
We first crawl the app reviews from Google Play Store submitted during August 2019 to January 2020, with the tool google-play-scraper\footnote{https://github.com/facundoolano/google-play-scraper}.
For each app, we then randomly sample around 550 reviews (about 1500 sentences) and label them for further experiments.
Table \ref{tab:label_data} elaborates the statistics of the experimental dataset in detail.
It contains 3,426 reviews and 8,788 sentences in total.



\begin{table}[tb]
\centering
\footnotesize
\caption{Experimental dataset.}
\label{tab:label_data}
\begin{tabular}{p{2cm}<{\centering}|p{2cm}<{\centering}|p{1.5cm}<{\centering}|p{1.5cm}<{\centering}}
\hline
\textbf{Category}                       & \textbf{App} & \textbf{\# Reviews} & \textbf{\# Sentences} \\ \hline
\multirow{2}{*}{\textbf{Social}}        & Instagram   & 582                & 1,402                 \\ \cline{2-4}
                                        & Snapchat     & 585                & 1,388                 \\ \hline
\multirow{2}{*}{\textbf{Communication}} & Gmail      & 586                & 1,525                 \\ \cline{2-4}
                                        & Yahoo Mail  & 542                & 1,511                 \\ \hline
\multirow{2}{*}{\textbf{Finance}}       & BPI Mobile   & 588                & 1,488                 \\ \cline{2-4}
                                        & Chase Mobile & 543                & 1,474                 \\
\cline{2-4} \hline\hline
\multicolumn{2}{c|}{\textbf{Overall}}                 & 3,426               & 8,788                 \\ \hline
\end{tabular}
\end{table}




Three authors then manually label the app reviews to serve as the ground-truth in verifying the performance of {\tool}. 
To guarantee the accuracy of the labeling outcomes, 
the first two authors firstly label the app reviews of an app independently, i.e., mark the beginning and ending position of the problematic features in each review sentence.
Second, the fourth author compares the labeling results, finds the difference, and organizes a face-to-face discussion among them three to determine the final label. 
All the six apps follow the same process.
For the first labeled app (\textit{Instagram}), the Cohen's Kappa is 0.78 between the two participants, while for the last labeled app (\textit{Chase Mobile}), the Cohen's Kappa is 0.86.
After two rounds of labeling, a common consensus is reached for every review sentence.


\subsection{Baselines}
\label{subsec:experiment_baseline}



\subsubsection{\textbf{Baselines for Problematic Feature Extraction}}
\
\newline
We select methods that can extract target phrases from app reviews as baselines for problematic feature extraction.
To the best of our knowledge, existing methods are mainly pattern-based, which can be classified into three types based on the techniques:
1) Part-of-Speech (PoS) Pattern: SAFE \cite{Johann2017SAFE} and PUMA \cite{VuPNN16}; 2) Dependency Parsing plus PoS Pattern: Caspar \cite{guo2020Caspar} and SUR-Miner \cite{Gu2015what}; 3) Pattern-based Filter plus Text Classification: KEFE \cite{WuDNN21}.
We select the representative method from each type as baselines, i.e., KEFE, Caspar, and SAFE.
In addition, since we model the feature extraction as an NER task, we also include BiLSTM-CRF \cite{HuangXY15}, a commonly-used technique in NER tasks, as a baseline.
We introduce four baselines in detail below:

\textbf{\textit{BiLSTM-CRF}} \cite{HuangXY15}:
A commonly-used algorithm in sequence tagging tasks such as NER.
Being a deep learning-based technique, it utilizes a BiLSTM to capture sentence semantics and a CRF layer to learn sentence-level tags.

\textbf{\textit{KEFE}} \cite{WuDNN21}:
A state-of-the-art approach for identifying key features from app reviews.
A key feature is referred as the features that are highly correlated to app ratings.
It firstly employs a pattern-based filter to obtain candidate phrases, and then a BERT-based classifier to identify the features.
Since its patterns are designed for Chinese language, we replace them with the patterns in SAFE \cite{Johann2017SAFE} to handle English reviews.

\textbf{\textit{Caspar}} \cite{guo2020Caspar}: 
A method for extracting and synthesizing user-reported mini stories regarding app problems from reviews.
We treat its first step, i.e., events extraction, as a baseline.
An event is referred as a phrase that is rooted in a verb and includes other attributes related to the verb.
It employed pattern-based and grammatical NLP techniques such as PoS tagging and dependency parsing on review sentences to address this task.
We use the implementation provided by the original paper\footnote{https://hguo5.github.io/Caspar}.


\textbf{\textit{SAFE}} \cite{Johann2017SAFE}:
A method for extracting feature-related phrases from reviews by 18 PoS patterns.
For example, the pattern \textit{Verb-Adjective-Noun} can extract features like ``delete old emails''. 
We implement all 18 patterns to extract the phrases based on the NLP toolkit NLTK\footnote{https://github.com/nltk/nltk}.



\subsubsection{\textbf{Baselines for problematic feature Clustering}}
\
\newline
We employ the following two baselines for problematic feature clustering, which are commonly used for mining topics of app reviews:

\textbf{\textit{K-Means}}:
It is a commonly-used clustering algorithm, and was employed to cluster the keywords of app reviews \cite{Vu2015mining}.
In this work, we first encode each problematic feature with TF-IDF \cite{SaltonG83} vectors, then run K-Means to cluster all problematic features into topics, following previous work \cite{Vu2015mining}.
We apply the implementation in the library scikit-learn\footnote{https://scikit-learn.org}.


\textbf{\textit{LDA}} \cite{BleiNJ01}:
It is a commonly-used topic clustering algorithm, and was utilized to group the app features \cite{GuzmanM14}.
In this work, we treat the extracted problematic features as documents and run LDA for topic modeling, following previous work \cite{GuzmanM14}.
We employ the implementation in the library Gensim\footnote{https://radimrehurek.com/gensim}.


\subsection{Experimental Setup}
\label{subsec:experiment_setup}



To answer RQ1, we conduct nested cross-validation \cite{Kohavi95} on the experimental dataset.
The inner loop is for selecting optimal hyper-parameters, which are used for evaluating performance in the outer loop.
In the outer loop, we randomly divide the dataset into ten folds, use nine of them for training,
and utilize the remaining one fold for testing the performance.
The process is repeated for ten times, and the average performance is treated as the final performance.
In the inner loop, we use eight folds for training and one fold for validation.
We run each baseline (see Section \ref{subsec:experiment_baseline}) to obtain its performance following the same experimental setup, and present the evaluation results on each app and on the overall dataset, respectively.

For RQ2, we design three variants of {\model} model to demonstrate the necessity of employed review attributes in our model architecture.
In detail, BERT-CRF, BERT+Cat-CRF, and BERT+SEN-CRF respectively represent the model without review attributes (i.e., only with text), the model without review description sentiment (i.e., with text and app category), and the model without app category (i.e., with text and review description sentiment).
We reuse other experimental setups as RQ1.





For RQ3, we manually build the ground-truth clustering results to evaluate the problematic feature clustering performance. 
The criteria for labeling are to group the features that represent the same functionality into one cluster.
More specifically, we randomly sample 100 problematic features for each app (600 in total) derived from the results of RQ1.
The two authors independently label these problematic features into clusters in the first round, where the Cohen’s Kappa between two authors reaches 0.81 (i.e., a satisfactory degree of agreement).
Then follow-up discussions are conducted until common consensus is reached.
Finally, the 600 problematic features were labeled into 20 groups.
Note that we do not specify the number of clusters in advance, because it is hard to decide the number in our context.
Our proposed clustering method does not need to specify this parameter as well.
Meanwhile, we run our approach and each baseline (see Section \ref{subsec:experiment_baseline}) to cluster these problematic features, and obtain each approach's clustering performance by comparing the predicted and ground-truth clustering results for each app and the overall dataset, respectively.


The experimental environment is a desktop computer equipped with an NVIDIA GeForce RTX 2060 GPU, intel core i7 CPU, 16GB RAM, running on Windows 10, and training the model takes about 2.5 hours for each fold nested cross-validation. 

\subsection{Evaluation Metrics}
\label{subsec:experiment_metrics}


\subsubsection{\textbf{Metrics for Problematic Feature Extraction}}
\
\newline
We use precision, recall, and F1-Score, which are commonly-used metrics, to evaluate the performance of {\tool} for problematic feature extraction.
We treat a problematic feature is correctly predicted if the predicted phrase from {\tool} for a review sentence of an app is the same as the ground-truth one.
Three metrics are computed as:
\begin{itemize}
    \item \textbf{Precision} is the ratio of the number of correctly predicted phrases to the total number of predicted phrases.
    \item \textbf{Recall} is the ratio of the number of correctly predicted phrases to the total number of ground-truth phrases.
    \item \textbf{F1-Score} is the harmonic mean of precision and recall. 
\end{itemize}

\subsubsection{\textbf{Metrics for Problematic Feature Clustering}}
\
\newline
Following previous work \cite{HuangCXLL18}, we use the commonly-used Adjusted Rand Index (ARI) \cite{Lawrence1985Comparing} and Normalized Mutual Information (NMI) \cite{NguyenEB10} to evaluate the clustering performance by comparing with the ground-truth clustering results.
Higher metric values indicate better clustering performance. 
For clarity, we denote $G$ as the ground-truth clustering result, and $C$ as the predicted clustering result. 

\textbf{Adjusted Rand Index (ARI)}:
It takes values in $[-1$, $1]$, reflecting the degree of overlap between the two clusters.
The raw Rand Index (RI) is computed by $RI=\frac{a+b}{\binom{n}{2}}$, where $a$ is the number of pairs that are assigned in the same cluster in $G$ and also in the same cluster in $C$, and $b$ is the number of pairs that are assigned in different clusters both in $G$ and $C$.
$\binom{n}{2}$ is the total number of unordered pairs in a set of $n$ phrases. 
The raw RI score is then ``adjusted for chance'' into the ARI score using the following scheme:
\begin{equation}
    ARI=\frac{RI-E\left ( RI \right )}{max\left ( RI \right )-E\left ( RI \right )}
\end{equation}
where $E\left ( RI \right )$ is the expected value of $RI$.
In this way, the ARI can be ensured to have a value close to 0.0 for random labeling independently of the number of clusters and samples.

\textbf{Normalized Mutual information (NMI)}:
It measures the similarity degree of the two sets of clustering results between 0 (no mutual information) and 1 (perfect correlation).
\begin{equation}
    NMI\left ( G,C \right )=\frac{MI\left ( G,C \right )}{\sqrt{H\left ( G \right )H\left ( C \right )}}
\end{equation}
where $H\left ( G \right )=-\sum _{i=1}^{\left | G \right |}P\left ( i \right )log\left ( P\left ( i \right ) \right )$ is the entropy of set $G$, and $P\left ( i \right )=\frac{G _i}{N}$ is the probability that a phrase 
picked randomly falls into cluster $G_i$.
The $MI\left ( G,C \right )$ is the mutual information of $G$ and $C$, i.e., $MI\left ( G,C \right )=\sum_{i=1}^{\left | G \right |}\sum_{j=1}^{\left | C \right |}P\left ( i,j \right )log\left ( \frac{P\left ( i,j \right )}{P\left ( i \right )P\left ( j \right )} \right )$.

\section{Results and Analysis}
\label{sec:results}

\subsection{Answering RQ1}
\label{subsec:results_RQ1.1}

The last column of Table \ref{tab:phrase} presents the performance of {\tool} in problematic feature extraction.
The overall precision, recall and F1 are 84.27\%, 85.06\% and 84.64\% respectively, which indicates that 84.27\% of problematic features extracted by {\tool} are correct, and 85.06\% problematic features are correctly extracted from the ground-truth ones.
The results confirm that 
our proposed approach can accurately extract the problematic features.

More specifically, {\tool} reaches the highest precision of 90.27\% on \textit{Gmail} and the highest recall of 87.37\% on \textit{Yahoo Mail}.
Its lowest precision is 79.18\% on \textit{Yahoo Mail} and the lowest recall is 84.15\% on \textit{Snapchat}.
We can see that even with its worst performance, an acceptable precision and recall can be achieved.

We then examine the extracted problematic features in detail, and find that there are indeed some observable patterns associated with the problematic features.
For example, users would use some negative words (e.g., ``cannot'', ``hardly'') or temporal conjunctions (e.g., ``as soon as'', ``when'') before mentioning the problematic features. 
This could probably explain why the pattern-based technique \cite{guo2020Caspar,Johann2017SAFE,gao2018INFAR} could work sometimes.
Taking the review in Figure \ref{fig:example} as an example, extracting the phrases after the negative word ``can't'' would obtain the correct phrase.
However, the pattern-based techniques highly rely on the manually defined patterns and have poor scalability in a different dataset.
Furthermore, 
there are many circumstances when the pattern-based approach can hardly work. 
For example, it is quite demanding to design patterns for the following review sentence: ``this update takes away my ability to view transactions'', where the problematic feature is ``view transaction''.
These circumstances further prove the advantages and flexibility of our approach.


\begin{table}[tb]
\centering
\footnotesize
\begin{threeparttable}[b]
\caption{Evaluation on problematic feature extraction (RQ1).}
\label{tab:phrase}
\begin{tabular}{p{1cm}<{\centering}|p{0.5cm}<{\centering}|p{0.9cm}<{\centering}|p{0.9cm}<{\centering}|p{0.9cm}<{\centering}|p{0.9cm}<{\centering}|p{0.9cm}<{\centering}}
\hline
\multicolumn{2}{c|}{\diagbox[height=0.9cm,width=2.2cm]{\textbf{App}}{\textbf{Metric}}{\textbf{Method}}} & \textbf{KEFE} & \textbf{Caspar} & \textbf{SAFE}    & \textbf{BiLSTM-CRF} & \textbf{SIRA}    \\ \hline
\multirow{3}{*}{\textbf{Instagram}}      & \textbf{P}       & 40.32\%       & 16.26\%         & 14.17\%          & 80.24\%             & \textbf{83.59\%} \\ 
                                         & \textbf{R}          & 60.76\%       & 10.49\%         & 70.61\%          & 71.79\%             & \textbf{85.70\%} \\ 
                                         & \textbf{F1}              & 48.29\%       & 12.46\%         & 23.55\%          & 75.58\%             & \textbf{84.53\%} \\ \hline
\multirow{3}{*}{\textbf{Snapchat}}       & \textbf{P}       & 42.08\%       & 18.87\%         & 12.95\%          & 78.49\%             & \textbf{82.63\%} \\ 
                                         & \textbf{R}          & 58.71\%       & 13.81\%         & 65.60\%          & 74.71\%             & \textbf{84.15\%} \\
                                         & \textbf{F1}              & 48.70\%       & 15.74\%         & 21.59\%          & 76.47\%             & \textbf{83.30\%} \\ \hline
\multirow{3}{*}{\textbf{Gmail}}          & \textbf{P}       & 53.79\%       & 25.60\%         & 22.25\%          & 87.58\%             & \textbf{90.27\%} \\
                                         & \textbf{R}          & 78.54\%       & 9.88\%          & \textbf{88.21\%} & 71.74\%             & 84.16\%          \\
                                         & \textbf{F1}              & 63.46\%       & 14.12\%         & 35.49\%          & 78.81\%             & \textbf{87.09\%} \\ \hline
\multirow{3}{*}{\textbf{\tabincell{c}{Yahoo\\ Mail}}}   & \textbf{P}       & 12.57\%       & 18.26\%         & 12.57\%          & 74.45\%             & \textbf{79.18\%} \\
                                         & \textbf{R}          & 70.10\%       & 11.85\%         & 70.10\%          & 74.69\%             & \textbf{87.37\%} \\
                                         & \textbf{F1}              & 21.25\%       & 14.19\%         & 21.25\%          & 74.26\%             & \textbf{83.00\%} \\ \hline
\multirow{3}{*}{\textbf{\tabincell{c}{BPI\\Mobile}}}     & \textbf{P}       & 41.92\%       & 20.98\%         & 18.22\%          & 82.58\%             & \textbf{87.37\%} \\
                                         & \textbf{R}          & 62.75\%       & 9.24\%          & 77.05\%          & 73.53\%             & \textbf{85.07\%} \\
                                         & \textbf{F1}              & 50.13\%       & 12.51\%         & 29.44\%          & 77.63\%             & \textbf{86.13\%} \\ \hline
\multirow{3}{*}{\textbf{\tabincell{c}{Chase\\Mobile}}} & \textbf{P}       & 36.98\%       & 17.53\%         & 12.17\%          & 77.23\%             & \textbf{80.32\%} \\
                                         & \textbf{R}          & 52.85\%       & 13.38\%         & 64.85\%          & 68.43\%             & \textbf{84.59\%} \\
                                         & \textbf{F1}              & 43.16\%       & 15.03\%         & 20.44\%          & 72.31\%             & \textbf{82.26\%} \\ \hline\hline
\multirow{3}{*}{\textbf{Overall}}        & \textbf{P}       & 42.79\%\tnote{$\ast$}       & 19.14\%\tnote{$\ast$}         & 15.51\%\tnote{$\ast$}          & 80.40\%             & \textbf{84.27\%} \\
                                         & \textbf{R}          & 63.50\%\tnote{$\ast$}       & 11.27\%\tnote{$\ast$}         & 73.94\%\tnote{$\ast\ast$}          & 72.48\%\tnote{$\ast$}             & \textbf{85.06\%} \\
                                         & \textbf{F1}              & 51.05\%\tnote{$\ast$}       & 14.13\%\tnote{$\ast$}         & 25.62\%\tnote{$\ast$}          & 76.15\%\tnote{$\ast$}             & \textbf{84.64\%} \\ \hline
\end{tabular}
\begin{tablenotes}
    \item Compared to SIRA, statistical significance $p-value < 0.05$ is denoted by $^{\ast\ast}$, and $p-value < 0.01$ is denoted by $^\ast$.
\end{tablenotes}
\end{threeparttable}
\vspace{-0.2in}
\end{table}
\begin{table*}[tb]
\centering
\footnotesize
\caption{Examples on extracted problematic features by different approaches (RQ1).}
\label{tab:extract_eg}
\begin{tabular}{l|l||l|l|l|l|l}
\hline
\textbf{\#} &   \multicolumn{1}{c||}{\textbf{Review}}   & \multicolumn{1}{c|}{\textbf{KEFE}} & \multicolumn{1}{c|}{\textbf{Caspar}} & \multicolumn{1}{c|}{\textbf{SAFE}} & \multicolumn{1}{c|}{\textbf{BiLSTM-CRF}} & \multicolumn{1}{c}{\textbf{{\tool}}} \\ \hline
\multicolumn{1}{c|}{\textbf{\# 1}} & \tabincell{l}{Keeps crashing \\when I try to take a picture of a check.}                                                 & take a picture  & \tabincell{l}{keeps crashing, I try to take \\ a picture of a check}         & \tabincell{l}{keeps crashing, \\take a picture}     & \tabincell{l}{take a picture \\ of a check}      & \tabincell{l}{take a picture \\ of a check}          \\ \hline
\multicolumn{1}{c|}{\textbf{\# 2}} & \tabincell{l}{When I try to view story of friend, the majority of\\ the time it get stuck on a wheel and never load.} & view story & \tabincell{l}{I try to view story of friend, \\the majority of the time it get \\ stuck on a wheel, \\never load}      & view story           & view story              & view story of friend               \\ \hline
\end{tabular}
\end{table*}

We also examine the bad cases where {\tool} fails to work. 
In some cases, {\tool} can extract the core nouns and verbs of the target phrase, but misses or additionally extracts some trivial words, especially some adverbs/adverbials before or after the core phrase.
For example, {\tool} might wrongly extract ``received emails for 10 days'' from ``I have not received emails for 10 days'', where the ground-truth phrase is ``received emails''.
Such results pull down the performance.
This could be improved by considering PoS patterns of words when vectorizing review sentences in future work.



\textbf{Comparison with baselines.}
Table \ref{tab:phrase} presents the performance of {\tool} and four baselines in extracting problematic features.
{\tool} outperforms all baselines on all metrics.
This indicates that these pattern-based baselines (i.e., KEFE, Caspar and SAFE) are far from effective in extracting problematic features, while the deep learning-based baseline (i.e., BiLSTM-CRF) is a bit worse than {\tool} because of the inferior semantic understanding and neglect of review attributes.
To further intuitively demonstrate the advantages of {\tool}, Table \ref{tab:extract_eg} presents two example reviews and the corresponding problematic features extracted by {\tool} and four baselines.


Among the three pattern-based baselines, SAFE achieves 15.51\% precision and 73.94\% recall.
This is because it defines 18 PoS patterns for feature-related phrases, and can retrieve a large number of possible problematic features (i.e., high recall).
For example, in the first example of Table \ref{tab:extract_eg}, SAFE would return two phrases.
By comparison, Caspar only extracts events from reviews containing temporal conjunctions and key phrases, including ``when'', ``every time'', which can hardly work well in this context. 
Taking the first review in Table \ref{tab:extract_eg} as an example, Caspar can only extract the two phrases/clauses.
KEFE achieves the promising performance, indicating that it can filter away many low-quality phrases with the BERT classifier; yet the classification is still conducted based on candidate phrases extracted by a pattern-based method, which limits its performance.
In the first example of Table \ref{tab:extract_eg}, KEFE can filter the wrong phrase ``keeps crashing'', but the reserved phrase ``take a picture'' is still not accurate enough due to the drawback of pattern-based candidate phrases.
BiLSTM-CRF can achieve promising performance but still not as accurate as our proposed {\tool}, e.g., ``view story'' in Table \ref{tab:extract_eg}.
{\tool} can be regarded as an improved version of BiLSTM-CRF, which employs BERT fine-tuning technique and two customized review attributes.
The features extracted by {\tool} is the superset of BiLSTM-CRF, which can be also reflected by the results in Table \ref{tab:phrase}.
{\tool} outperforms BiLSTM-CRF in both recall and precision, indicating that {\tool} can extract features more accurately and retrieve more problematic features.

\subsection{Answering RQ2}
\label{subsec:results_RQ1.3}

\begin{table}[tb]
\centering
\footnotesize
\begin{threeparttable}[b]
\caption{Ablation experiment on attributes (RQ2).}
\label{tab:ablation}
\begin{tabular}{p{1cm}<{\centering}|p{1cm}<{\centering}|p{1cm}<{\centering}|p{1cm}<{\centering}|p{1cm}<{\centering}|p{1cm}<{\centering}}
\hline
\multicolumn{2}{c|}{\diagbox[height=0.8cm]{\textbf{App}}{\textbf{Metric}}{\textbf{Method}}}                           & \textbf{\tabincell{c}{BERT\\-CRF}} & \textbf{\tabincell{c}{BERT\\+CAT\\-CRF}}              & \textbf{\tabincell{c}{BERT\\+SEN\\-CRF}}              & \textbf{\tabincell{c}{BERT\\+Attr\\-CRF}}    \\ \hline
\multirow{3}{*}{\textbf{Instagram}}    & \textbf{P}       & 82.46\%          & \textbf{84.08\%}   & 83.78\%            & 83.59\%          \\
                                       & \textbf{R}          & 80.39\%          & 85.60\%            & 85.50\%            & \textbf{85.70\%} \\
                                       & \textbf{F1}              & 81.34\%          & \textbf{84.73\%}   & 84.56\%            & 84.53\%          \\ \hline
\multirow{3}{*}{\textbf{Snapchat}}     & \textbf{P}       & \textbf{84.58\%} & 83.82\%            & 83.38\%            & 82.63\%          \\
                                       & \textbf{R}          & 81.49\%          & 83.31\%            & \textbf{85.31\%}   & 84.15\%          \\
                                       & \textbf{F1}              & 82.89\%          & 83.48\%            & \textbf{84.23\%}   & 83.30\%          \\ \hline
\multirow{3}{*}{\textbf{Gmail}}        & \textbf{P}       & 88.33\%          & 89.30\%            & \textbf{90.59\%}   & 90.27\%          \\
                                       & \textbf{R}          & 78.37\%          & 83.43\%            & 83.50\%            & \textbf{84.16\%} \\
                                       & \textbf{F1}              & 82.99\%          & 86.16\%            & 86.86\%            & \textbf{87.09\%} \\ \hline
\multirow{3}{*}{\textbf{\tabincell{c}{Yahoo\\Mail}}}   & \textbf{P}       & 75.92\%          & 76.67\%            & 78.23\%            & \textbf{79.18\%} \\
                                       & \textbf{R}          & 83.72\%          & 83.72\%            & 86.09\%            & \textbf{87.37\%} \\
                                       & \textbf{F1}              & 79.54\%          & 79.94\%            & 81.86\%            & \textbf{83.00\%} \\ \hline
\multirow{3}{*}{\textbf{\tabincell{c}{BPI\\Mobile}}}   & \textbf{P}       & 84.87\%          & 85.92\%            & 85.52\%            & \textbf{87.37\%} \\
                                       & \textbf{R}          & 78.09\%          & 84.94\%            & 82.60\%            & \textbf{85.07\%} \\
                                       & \textbf{F1}              & 81.25\%          & 85.32\%            & 83.96\%            & \textbf{86.13\%} \\ \hline
\multirow{3}{*}{\textbf{\tabincell{c}{Chase\\Mobile}}} & \textbf{P}       & 78.24\%          & 80.26\%            & 80.05\%            & \textbf{80.32\%} \\
                                       & \textbf{R}          & 77.59\%          & 82.19\%            & 83.74\%            & \textbf{84.59\%} \\
                                       & \textbf{F1}              & 77.73\%          & 81.11\%            & 81.76\%            & \textbf{82.26\%} \\ \hline\hline
\multirow{3}{*}{\textbf{Overall}}      & \textbf{P}       & 82.59\%          & 83.73\%            & 83.95\%            & \textbf{84.27\%} \\
                                       & \textbf{R}          & 79.69\%         & 83.88\%\tnote{$\ast$}           & 84.31\%\tnote{$\ast$}           & \textbf{85.06\%}\tnote{$\ast$}\\
                                       & \textbf{F1}              & 81.10\%          & 83.78\%\tnote{$\ast\ast$}           & 84.10\%\tnote{$\ast\ast$}           & \textbf{84.64\%}\tnote{$\ast$} \\ \hline
\end{tabular}
\begin{tablenotes}
    \item Compared to BERT-CRF, statistical significance $p-value < 0.05$ is denoted by $^{\ast\ast}$, and $p-value < 0.01$ is denoted by $^\ast$.
\end{tablenotes}
\end{threeparttable}
\end{table}

Table \ref{tab:ablation} presents the performance of {\tool} and its three variants, respectively.
The overall performance of {\tool} is higher than all the three variants.
Compared with the base BERT-CRF model, adding the app category and the 
sentiment attributes noticeably increase the precision (2.03\%) and recall (6.74\%).
This indicates that the 
two attributes are  helpful in identifying the problematic features.
For the performance on each app, adding the two attributes (i.e., {\model}) obtains the best performance on most apps, and adding one of the two attributes (i.e., BERT+CAT-CRF or BERT+SEN-CRF) occasionally achieves the best performances on some apps (e.g., BERT+SEN-CRF on \textit{Snapchat}).
Moreover, even the performance of the base BERT-CRF model outperforms the best baseline in RQ1 (i.e., BiLSTM-CRF), which verifies the advantage of our model design.


Among the two added review attributes, the review description sentiment attribute contributes slightly more to performance improvement (1.64\% in precision and 5.80\% in recall) than the app category attribute (1.38\% in precision and 5.26\% in recall).
Furthermore, we also observe that the contribution of these two attributes overlaps to some extent, i.e., the increased performance by each attribute is not simply added up to the performance of the whole model.
This is reasonable considering the fact that words expressing the user sentiment could be encoded semantically in the textual descriptions and captured by the BERT model. 
Nevertheless, the overall performance achieved by adding both of the attributes is the highest, further indicating the necessity of our model design. 


\subsection{Answering RQ3}
\label{subsec:results_RQ2.1}

\begin{table}[tb]
\centering
\footnotesize
\caption{Evaluation on problematic feature clustering (RQ3).}
\label{tab:cluster_evl}

\begin{tabular}{p{1.5cm}<{\centering}|p{1.5cm}<{\centering}|p{1.2cm}<{\centering}|p{1.2cm}<{\centering}|p{1.2cm}<{\centering}}
\hline
\multicolumn{2}{c|}{\diagbox[width=3.7cm]{\textbf{App}}{\textbf{Metric}}{\textbf{Method}}}                                 & \textbf{LDA} & \textbf{K-Means} & \textbf{{\tool}}                \\ \hline
\multirow{2}{*}{\textbf{Instagram}}    & \textbf{ARI} & 0.10          & \textbf{0.30}    & 0.29          \\
                                       & \textbf{NMI} & 0.72          & 0.78             & \textbf{0.84} \\ \hline
\multirow{2}{*}{\textbf{Snapchat}}     & \textbf{ARI} & 0.19          & 0.13             & \textbf{0.32} \\
                                       & \textbf{NMI} & 0.80          & 0.72             & \textbf{0.85} \\ \hline
\multirow{2}{*}{\textbf{Gmail}}        & \textbf{ARI} & 0.18          & 0.07             & \textbf{0.45} \\
                                       & \textbf{NMI} & 0.73          & 0.58             & \textbf{0.82} \\ \hline
\multirow{2}{*}{\textbf{Yahoo Mail}}   & \textbf{ARI} & 0.42          & \textbf{0.47}    & 0.41          \\
                                       & \textbf{NMI} & 0.81          & \textbf{0.83}    & 0.82          \\ \hline
\multirow{2}{*}{\textbf{BPI Mobile}}   & \textbf{ARI} & 0.44          & 0.10             & \textbf{0.59} \\
                                       & \textbf{NMI} & 0.83          & 0.58             & \textbf{0.89} \\ \hline
\multirow{2}{*}{\textbf{Chase Mobile}} & \textbf{ARI} & \textbf{0.38} & 0.21             & 0.26          \\
                                       & \textbf{NMI} & 0.81          & 0.79             & \textbf{0.82} \\ \hline\hline
\multirow{2}{*}{\textbf{Overall}}      & \textbf{ARI} & 0.21          & 0.14             & \textbf{0.38} \\
                                       & \textbf{NMI} & 0.57          & 0.62             & \textbf{0.77} \\ \hline
\end{tabular}
\end{table}

Table \ref{tab:cluster_evl} presents the performance of {\tool} in clustering problematic features, as well as the two baselines.
{\tool} outperforms the two baselines on the overall performance, where ARI and NMI reach 0.38 and 0.77, respectively, which is higher than that of LDA (0.21 and 0.57) and K-Means (0.14 and 0.62). 

Furthermore, the improvement of {\tool} on ARI is greater than the improvement on NMI. 
ARI is a pair-wise metric, which is more sensitive when two phrases that should belong to the same cluster are wrongly assigned into different clusters, or when two phrases which should belong to different clusters are wrongly placed into the same cluster.
The ARI results we obtained indicate that {\tool} can effectively avoid generating new clusters or breaking up the original clusters. NMI is an entropy-based metric, which mainly focuses on the changes of two distributions based on information entropy theory.
The NMI results we obtained indicate that the distribution of the entire cluster (e.g., the number of problematic features in each cluster) derived from {\tool} are closer to the ground-truth.

The baseline approaches use the word statistics or co-occurrence relations to cluster the problematic features. 
The performance of our proposed graph-based clustering method indicates that it can better understand the semantic relations among problematic features. 
\section{Where the Apps Frustrate Users - An Empirical Study with {\tool}}
\label{sec:crowdtesting}




\begin{figure*}[tb]
	\centering
  	\subfigure[Social]{
		\includegraphics[width=0.66\columnwidth]{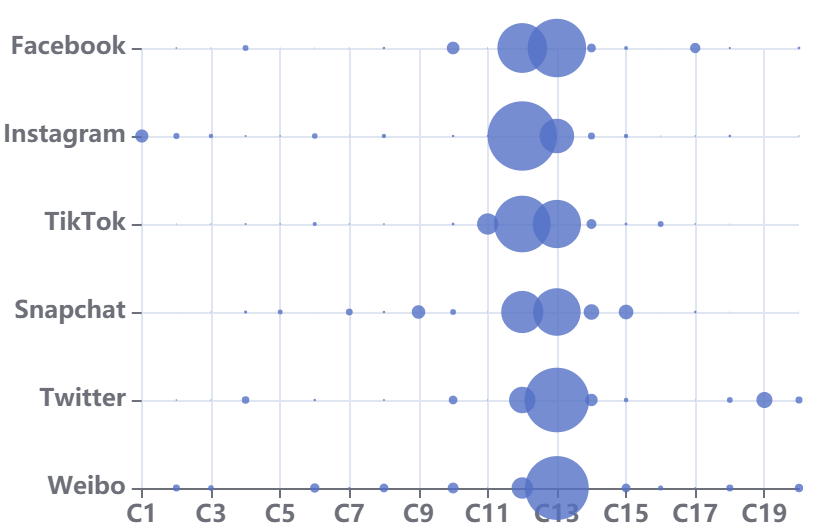}
		\label{subfig:cluster_social}
		}
	\subfigure[Communication]{
		\includegraphics[width=0.66\columnwidth]{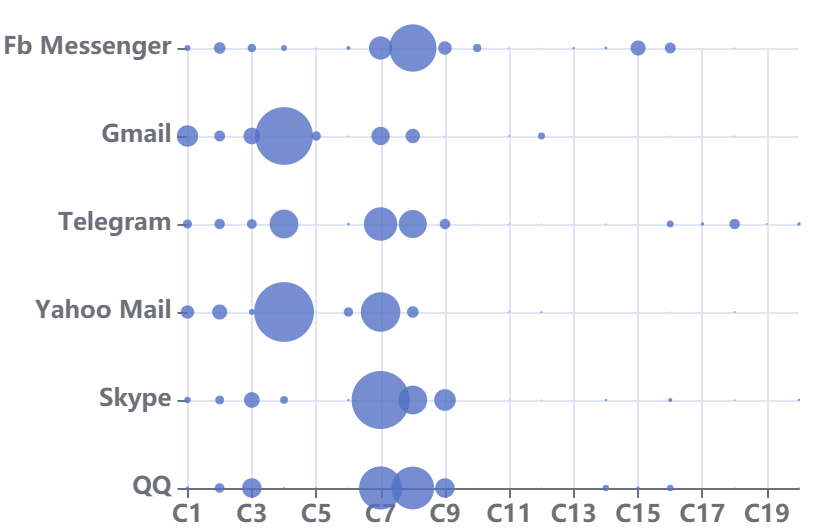}
		\label{subfig:cluster_commu}
		}
	\subfigure[Finance]{
		\includegraphics[width=0.66\columnwidth]{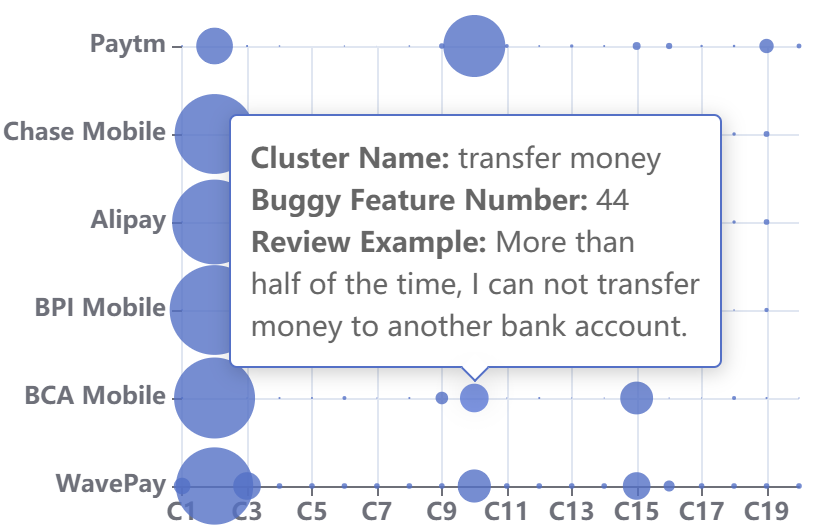}
		\label{subfig:cluster_finance}
		}
	\caption{The distribution of problematic features of different categories.}
    \label{fig:cluster_stats}
\end{figure*}
\begin{table}[tb]
\centering
\footnotesize
\caption{Experimental dataset for investigating ``where the apps frustrating users''.}
\label{tab:unlabel_data}
\begin{tabular}{p{2cm}<{\centering}|p{1.5cm}<{\centering}|p{1.5cm}<{\centering}|p{1.5cm}<{\centering}}
\hline
\textbf{Category}                       & \textbf{App}       & \textbf{\# Reviews} & \textbf{\# Sentences} \\ \hline
\multirow{5}{*}{\textbf{Social}}        & Facebook           & 64,559              & 147,156               \\ \cline{2-4} 
                                        & Instagram          & 63,124              & 153,852                \\ \cline{2-4} 
                                        & TikTok             & 61,178              & 104,094               \\ \cline{2-4} 
                                        & Snapchat           & 18,268              & 41,278               \\ \cline{2-4} 
                                        & Twitter            & 15,583              & 36,386
                                        \\ \cline{2-4}
                                        & Sina Weibo              & 10,772              & 37,372
                                        \\ \hline
\multirow{5}{*}{\textbf{Communication}} & \tabincell{c}{Facebook\\-Messenger} & 27,121              & 59,303                \\ \cline{2-4} 
                                        & Gmail              & 9,655               & 24,520                \\ \cline{2-4} 
                                        & Telegram           & 7,704               & 17,672                 \\ \cline{2-4} 
                                        & Yahoo Mail         & 7,090               & 20,124                \\ \cline{2-4} 
                                        & Skype              & 3,266               & 8,139               
                                        \\ \cline{2-4}
                                        & Tencent QQ                 & 3,194               & 7,326
                                        \\ \hline
\multirow{5}{*}{\textbf{Finance}}       & Paytm              & 18,316              & 47,836                 
                                        \\ \cline{2-4} 
                                        & Chase Mobile       & 3,732               & 9,952                
                                        \\ \cline{2-4}
                                        & Alipay             & 3,153               & 9,359
                                        \\ \cline{2-4} 
                                        & BPI Mobile         & 1,375               & 3,638              
                                        \\ \cline{2-4} 
                                        & BCA Mobile         & 386                & 960                  
                                        \\ \cline{2-4} 
                                        & WavePay            & 58                 & 124                
                                        \\ \hline\hline
\multicolumn{2}{c|}{\textbf{Overall}}                        & 318,534            & 729,091
                                        \\ \hline
\end{tabular}
\end{table}




This section describes a large-scale empirical study with {\tool} on popular apps. First, we apply {\tool} to 18 apps of three categories (6 in each category) to demonstrate: 1) how {\tool} can be utilized in real-world practice; 2) the distribution of problematic features across these popular apps.
We also select 3 apps (1 in each category) and conduct a user survey to verify the usefulness of {\tool}.

\textbf{{\tool} in the Large.}
We crawl the app reviews of 18 apps from three categories (6 in each category) submitted during February 2020 to December 2020 (note that this is different from the time period in Section \ref{subsec:experiment_data_prepare}).
Table \ref{tab:unlabel_data} lists the statistics of this dataset, which contains 318,534 reviews and 729,091 sentences.
We run {\tool} on this large-scale dataset to obtain the visualization of the clustered problematic features (see Section \ref{subsec:approach_visualization}).
In total, we obtain 113 clusters for social apps, 78 clusters for communication apps and 90 clusters for finance apps.
Figure \ref{fig:cluster_stats} presents the visualization results of clusters for each category with the bubble size denoting the ratio of corresponding problematic features.
For clarity, we only present the clusters whose number of problematic features is in top 20, by the order of cluster id.
Table \ref{tab:keyword} shows the name of each cluster in Figure \ref{fig:cluster_stats}.
The following observations can be obtained.


First, our visualization can provide a summarized view of the problematic features for each app and the comparison across apps. 
This enables the developers to acquire where the app is prone to problems, and where other apps are also likely to have issues, with a single glance. 
One can also derive the detailed content of each cluster, and example app reviews of the cluster by hovering the cursor over the bubble in the figure (see examples in Figure \ref{subfig:cluster_finance}). 


\begin{table}[tb]
\centering
\footnotesize
\caption{Cluster name (i.e., representative problematic feature) of each cluster in Figure \ref{fig:cluster_stats}.}
\label{tab:keyword}
\begin{tabular}{c|c|c|c}
\hline
\textbf{\#}  & \textbf{Social}       & \textbf{Communication}        & \textbf{Finance}       \\ \hline
\textbf{C1}  & the reel option       & delete email                  & send message           \\ \hline
\textbf{C2}  & like a post           & open app                      & log in                 \\ \hline
\textbf{C3}  & search option         & receive notification          & receive otp code       \\ \hline
\textbf{C4}  & load tweet            & send and receive email        & load the page          \\ \hline
\textbf{C5}  & use filter            & dark mode                     & check deposit          \\ \hline
\textbf{C6}  & follow people         & load inbox                    & get notification       \\ \hline
\textbf{C7}  & the front camera      & sign into account             & use finger print       \\ \hline
\textbf{C8}  & click on photo        & send picture and video        & click button           \\ \hline
\textbf{C9}  & send snap             & video call                    & do transaction         \\ \hline
\textbf{C10} & receive notification  & see story                     & transfer money         \\ \hline
\textbf{C11} & get live option       & click on call button          & get cash back          \\ \hline
\textbf{C12} & post story            & sync account                  & scan qr code           \\ \hline
\textbf{C13} & access account        & \tabincell{c}{change the emoji\\and nickname} & \tabincell{c}{recharge mobile\\number} \\ \hline
\textbf{C14} & open snap             & share photo                   & change phone number    \\ \hline
\textbf{C15} & send message          & register user                 & open passbook          \\ \hline
\textbf{C16} & watch video           & chat with friend              & book ticket            \\ \hline
\textbf{C17} & dark mode             & get otp for login             & select option          \\ \hline
\textbf{C18} & scroll the feed       & receive verification code     & check balance          \\ \hline
\textbf{C19} & retrieve tweet        & quiz bot                      & make payment           \\ \hline
\textbf{C20} & get verification code & change phone number           & receive the refund     \\ \hline
\end{tabular}
\end{table}

Second, different apps can share similar problematic features, which can facilitate app 
testing and refine the testing techniques. 
Take Figure \ref{subfig:cluster_social} as an example, although the problematic features are observed distributing differently across apps, all the six investigated apps would have a noticeable number of problematic features in certain clusters (i.e., \textit{C12. post story} and \textit{C13. access account}).
These information can warn the developers of similar apps to notice potential problems, especially which have not yet been reported or only mentioned in a few reviews.
Further, developers can leverage reviews from similar apps for quality assurance activities, rather than only focus on the limited set of reviews of its own app.
This is especially the case for the less popular apps which only have few reviews regarding app problems.

Third, different apps can have their unique problematic features and problematic feature distributions, which further indicates the necessity of review mining and analysis in a fine-grained way.
For example, from Figure \ref{subfig:cluster_commu}, we can see that, based on the user reported problems, 63\% reviews of the \textit{Facebook Messenger} app relate with feature \textit{C8. send picture and video}.
By comparison, its competitor \textit{Gmail} app is mainly prone to bugs for quite different feature \textit{C4. send and receive email}.
In addition, for its another competitor \textit{Telegram} app, the problematic features are distributed more evenly, i.e., the number of user submitted reviews do not exert big difference across \textit{C4}, \textit{C7} and \textit{C8}, and the largest cluster (i.e., \textit{C7. sign into account}) occupies a mere of 33\% reviews. 
From these insights provided by our approach, the developers can obtain a clear understanding of an app about the features that are prone to problems, so as to arrange the follow-up problem solving and allocate the testing activity for subsequent versions.
More than that, these information can also assist the developers in the competitive analysis of apps, e.g., acquire the weakness of their app compared with similar apps.


Furthermore, a series of attempts can be made to refine the app testing techniques. 
For example, one can recommend problematic features to similar apps in order to prioritize the testing effort, or recommend related descriptions (mined from app reviews) to similar apps to help bug detection. 
In addition, the automated graphical user interface (GUI) testing techniques can be customized and the testing contents can be prioritized. 
Current automated GUI testing tools tend to dynamically explore different pages of a mobile app through random actions (e.g., clicking, scrolling, etc) to trigger the crash or explicit exceptions \cite{li2017droidbot}.
If one could know the detailed problematic features of other similar apps in advance, the explored pages can be re-ranked so that the bug-prone features can be explored earlier to facilitate the bugs being revealed earlier.
We will further explore problematic features based app testing in our future work.
\begin{figure}[tb]
\centering
\includegraphics[width=0.9\columnwidth]{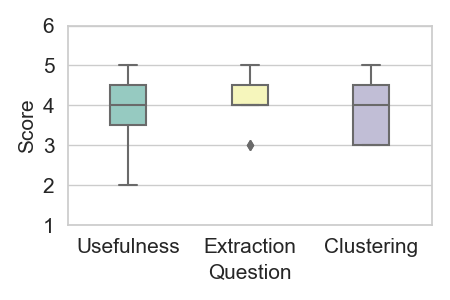}
\caption{Feedback of user study.}
\label{fig:feedback}
\end{figure}


\textbf{A User Survey.}
{In order to assess the usefulness of {\tool}, we conduct a user survey on three popular apps: \textit{Weibo}, \textit{QQ} and \textit{Alipay}.} 
We invite 15 respondents (5 from each company) in total, including 2 product managers, 5 requirement analysts, and 8 developers, who are familiar with the app reviews of their own company.
More specifically, we conduct {\tool} on the reviews obtained in the first week of May 2021, which contains 177 reviews from \textit{Weibo}, 149 from \textit{QQ}, and 177 from \textit{Alipay} after preprocessing.
Each respondent examines the extracted problematic features, clusters and visualization results obtained by {\tool}, and answer the following three questions:
1) (Usefulness) Can {\tool} help understand user requirements from app reviews?
2) (Extraction) Can {\tool} extracted problematic features accurately?
3) (Clustering) Can {\tool} cluster problematic features accurately?
We provide five options for each question from 1 (strongly disagree) to 5 (strongly agree). 
The first question concerns the usefulness of {\tool}, i.e., whether {\tool} can save effort for analyzing large-scale app reviews.
The last two questions concern the performance of {\tool} on problematic feature extraction and clustering respectively, when analyzing app reviews in real-world practice.

Figure \ref{fig:feedback} shows the box plot statistics of respondents' feedback.
There are respectively 11, 13 and 10 (out of 15) respondents give the score over 3 for Q1, Q2, and Q3.
Most of them (over 73\%) are satisfied (score over 3) with the usefulness of {\tool}, and think {\tool} can help them obtain a fine-grained understanding on problematic features.
The average score of Q1, Q2, and Q3 are 3.93, 4.13, and 3.93 respectively.
Besides, three of them heard about or tried existing review analysis tools such as INFAR \cite{gao2018INFAR} and SUR-Miner \cite{Gu2015what}, and they admit the advantages of {\tool} as its extracted features and derived clusters are finer-grained and more meaningful.
We also interviewed the respondents about the possible enhancement of {\tool}.
They said there were still some cases where {\tool} doesn't work well, such as some extracted phrases contain two or more features, which leads to poor performance of clustering.
This can be solved in future work by exploring the patterns of such tangled features and deconstructing them into separate ones.
In addition, we received some suggestions from developers for better visualizations (e.g., supporting interactive visual analytics).

\section{Discussion}
\label{sec_discussion}

\textbf{Advantage Over Topic Discovery Approaches.}
There are several previous approaches which involve topic discovery  \cite{Vu2015mining,Gu2015what,DiSorbo2016what,gao2018INFAR,VuPNN16}. 
Yet, their discovered topics are more coarse-grained than our proposed approach. 
For example, based on 95 mobile apps like \textit{Facebook} and \textit{Twitter} from Google Play,  MARK \cite{Vu2015mining} can only discover such topics as \textit{crash}, \textit{compatibility}, and \textit{connection}, and PUMA \cite{VuPNN16} generates topics like \textit{battery consumption.}
Similarly, SUR-Miner \cite{Gu2015what} generates topics such as \textit{predictions}, \textit{auto-correct}, and \textit{words}.
SURF \cite{DiSorbo2016what} can discover topics such as \textit{GUI}, \textit{app}, and \textit{company}, while INFAR \cite{gao2018INFAR} can generate topics like \textit{update}, \textit{radar}, \textit{download}.
With these discovered topics, the developers can acquire a general view about the problems the app undergoes, yet could not get a clear understanding about where it is wrong.
By comparison, as demonstrated in Figure \ref{fig:cluster_stats} and Table \ref{tab:keyword}, our proposed approach can generate more finer-grained topics as \textit{open message, get cash back}, 
which helps developers achieve a deeper and more accurate understanding about where the app is wrong. 

\textbf{Threats to Validity.}
The \textbf{\textit{external threats}} concern the generality of the proposed approach.
We train and evaluate {\tool} on the dataset consisting of six apps from three categories.
The selected apps and their belonging categories are all the commonly-used ones with rich reviews in practice, which relatively reduces this threat.
In addition, we demonstrate the usage of {\tool} on a much bigger dataset derived from 18 apps.
The results are promising, which verifies its generality further.
Regarding \textbf{\textit{internal threats}}, {\tool} is a pipeline method, where the problematic feature clustering depends on the accuracy of extracting problematic features.
Since we have seen a relatively high performance of {\tool} on problematic feature extraction, we believe {\tool} can alleviate the error accumulation to some extent.
In addition, we reuse the source code from the original paper (i.e., for \textit{Caspar} and \textit{KEFE}), or the open source implementation (i.e., for \textit{SAFE}, \textit{K-Means}, and \textit{LDA}) for the baselines, which help ensure the accuracy of the experiments.
The \textbf{\textit{construct validity}} of this study mainly questions the evaluation metrics.
We utilize precision, recall and F1-Score to evaluate the performance of problematic feature extraction.
We consider that a problematic feature is correctly extracted when it is the same as the ground-truth, which is a rather strict measure.
The metrics used to evaluate clustering results are also commonly used \cite{HuangCXLL18}.


\section{Conclusion}
\label{sec:conclusion}

To help acquire a concrete understanding about where the app is frustrating the users, this paper proposes a semantic-aware, fine-grained app review analysis approach {\tool}, which can extract, cluster, and visualize the problematic features of app reviews.
{\tool} designs a novel  {\model} model to extract fine-grained problematic features, and employs a graph-based clustering method to cluster them.
We evaluate {\tool} on 3,426 reviews from six apps, and the results confirm the effectiveness of the proposed approach. 
We further conduct an empirical study on 318,534 reviews from 18 popular apps to explore its potential application 
and usefulness in real-world practice.
Our source code and experimental data are publicly available at: \url{https://github.com/MeloFancy/SIRA}.


\section*{Acknowledgments}
This work is supported by the National Key Research and Development Program of China under grant No.2018YFB1403400, the National Natural Science Foundation of China under grant No.62072442, the Youth Innovation Promotion Association Chinese Academy of Sciences, and Australian Research Council Discovery Project DP220103044.

\balance
\bibliographystyle{ACM-Reference-Format}
\bibliography{reference}


\end{document}